\documentclass[aps,prd,preprintnumbers,nofootinbib,twocolumn]{revtex4}
\usepackage{bm}
\usepackage{latexsym}
\usepackage{dcolumn}
\usepackage{amsmath,amsfonts,amssymb}
\usepackage{graphicx,epsfig}
\usepackage{psfrag}
\usepackage{amsthm}
\usepackage{color}

\usepackage{bm}
\usepackage{latexsym}
\usepackage{dcolumn}
\usepackage{amsmath,amsfonts,amssymb}
\usepackage{graphicx,epsfig}
\usepackage{psfrag}
\usepackage{amsthm}
\usepackage{color}

\usepackage{nccmath}
\usepackage{moresize}
\usepackage{enumerate}
\usepackage[hang,flushmargin]{footmisc}
\usepackage[titletoc,toc]{appendix}
\usepackage{lipsum}
\usepackage{subcaption}
\usepackage{epstopdf} 
\usepackage{hyperref}
\usepackage{rotating}
\usepackage{multirow}
\usepackage{mathtools}

\usepackage{stackengine}
\usepackage{scalerel}

\newcommand{\be}{\begin{equation}}
\newcommand{\ee}{\end{equation}}
\newcommand{\bea}{\begin{eqnarray}}
\newcommand{\eea}{\end{eqnarray}}
\newcommand{\bse}{\begin{subequations}}
\newcommand{\ese}{\end{subequations}}
\newcommand{\bce}{\begin{center}}
\newcommand{\ece}{\end{center}}
\newcommand{\bfg}{\begin{figure}}
\newcommand{\efg}{\end{figure}}
\newcommand{\bit}{\begin{itemize}}
\newcommand{\eit}{\end{itemize}}
\newcommand{\bed}{\begin{description}}
\newcommand{\eed}{\end{description}}
\newcommand{\ben}{\begin{enumerate}}
\newcommand{\een}{\end{enumerate}}
\newcommand{\nn}{\nonumber}

\newcommand{\pa}{\partial}
\newcommand{\fr}{\frac}

\newcommand{\no}{\noindent}

\def\le {\left}
\def\ri {\right}
\def\r  {\rho}

\newcommand{\cR}{\mathcal R}
\newcommand{\cS}{\mathcal S}

\newcommand{\vx}{\vec{\pmb x}}

\newcommand{\vJ}{\vec{\pmb J}}

\newcommand{\bdm}{\begin{displaymath}}
\newcommand{\edm}{\end{displaymath}}

\begin{document}


\title{On the quantum origin of potentials}

\author{Saurya Das} \email[email: ]{saurya.das@uleth.ca}
%


\affiliation{Theoretical Physics Group,
Department of Physics and Astronomy,
University of Lethbridge, 4401 University Drive,
Lethbridge, Alberta T1K 3M4, Canada}

\author{Sourav Sur}
\email[email: ]{sourav.sur@gmail.com}

\affiliation{Department of Physics and Astrophysics, University of Delhi, New Delhi 110007, India}

\begin{abstract}
The dynamics of a quantum particle is governed by its wavefunction, which 
in turn is determined by the classical potential to which it is subjected.
However the wavefunction itself induces a quantum potential, the particle 
`sees' the sum of the classical and quantum potentials, and there is no way 
to separate the two. Therefore in principle, part or whole of an observed 
potential may be attributable to a quantum potential. We examine this 
possibility and discuss implications. 

\end{abstract}

\maketitle

The standard paradigm of quantum mechanics, non-relativistic or relativistic, consists of 
solving a suitable wave equation (such as Schr\"odinger, Klein-Gordon or Dirac) with an appropriate potential, for 
normalizable wavefunctions.
%
However, there is a lesser known result that the wavefunction of a quantum 
particle itself generates a `quantum potential' and the dynamics of the particle 
is governed by the sum of the classical and quantum potentials and not by them 
individually. Therefore, part or all of one can be replaced by the other, and 
in principle it is possible that particles are intrinsically in a flat spacetime 
without any potentials and what is normally perceived as a classical potential, 
for e.g. a harmonic oscillator, Coulomb or the gravitational potential, is in 
reality, a quantum potential generated by the wavefunction of the particle. 

To explore this possibility, for simplicity, let us start with the Schr\"odinger 
equation (generalization to relativistic wave equations should be straightforward)
\be
-\, \fr{\hbar^2}{2m}\, \nabla^2 \Psi +\, V \Psi =\, i\, \frac{\partial \Psi}
{\partial t}~,
\label{SE1}
\ee
with the wavefunction written as
\be 
\Psi (\vx,t) =\, \cR (\vx, t) \, e^{i \,\cS(\vx,t)} \, \quad \le[\cR, \cS \in \mathbb{R}\ri] \,.
\label{wf1}
\ee 
Substituting this back in Eq.\,(\ref{SE1}), and extracting the real and imaginary 
parts, yields two equations 
\cite{bohm}
\bea 
&& \vec \nabla \cdot \vJ +\, \fr{\pa\r}{\pa t} =\, 0 \,, \label{consv-eq} \\
&& m \, \fr{d\vec v}{dt} =\, - \, \vec\nabla \le(V +\, V_Q\ri) = -\vec\nabla 
V_{tot}\,,  \label{Newt-eq}
\eea 
where $\rho = |\Psi|^2 \,,~ \vJ = \frac{\hbar}{2mi} \left[\Psi^\star 
\vec\nabla \Psi - \Psi \vec\nabla \Psi^\star\right]$ are respectively the 
probability density and the probability current density, 
%
%
$\vec v := (\hbar/m)\,\vec\nabla \cS \,$ is the `velocity field', 

\be \label{QP}
V_Q =\, -\, \fr{\hbar^2}{2m} \fr{\nabla^2 \cR} \cR \,, 
\ee 
is the `quantum potential', and 
\be
V_{tot} \equiv V + V_Q~.
\label{vtot}
\ee
%
Eq.\,(\ref{consv-eq}) is simply the standard conservation equation for $\r$ 
and $\vJ$. 
Eq.\,(\ref{Newt-eq}), on the other hand, shows that the quantum dynamics of a 
particle of mass $m$ is still governed by the classical Newton's equation, 
albeit with the classical potential $V$ augmented by the wavefunction dependent
`quantum potential'. 
For instance, while a stream of free electrons, governed just by a vanishing 
classical potential ($V = 0$), follows straight line trajectories, that 
governed by $V_{tot}=V + V_Q$, for an appropriate choice of 
$\Psi(\vx,t)$ reproduces the familiar interference patterns 
\cite{fringes}.
Thus, we see that a quantum particle subjected to $V_{tot}$ has no way 
of figuring out what part of it is `classical' and what part `quantum', and  
a part (or all) of the classical potential may in fact be the quantum potential
generated by a suitable wavefunction and vice-versa. Going one step further, 
one can even say that observed potentials and wavefunctions are interchangeable 
and it may be a matter of choice as to which one is considered more fundamental. 

Before proceeding further, we present an even lesser known
result which follows from Eqs.\,(\ref{SE1}) and (\ref{QP}), namely, 
that for a {\em stationary} state, of energy $E$ and described by the 
wavefunction $\, \Psi (\vx,t) = \psi(\vx)\, e^{-i E t/\hbar}\,$, one 
finds\footnote{
%
The easiest way is to see this is to recognize that 
$\nabla^2 {\cal R}/{\cal R} = \nabla^2 {\psi}/{\psi}$
for stationary states, and substituting for $\nabla^2 \psi$ using the 
time-independent Schr\"odinger equation 
%
$-\, \fr{\hbar^2}{2m}\, \nabla^2 \psi +\, V \psi =\, E \psi \, $
%
%
in Eq.(\ref{QP}).
}
\be \label{QP1}
V_Q =
\, -\, V +\, E \, .
\ee
%
The result is remarkable for two reasons.
First, it shows that 
the $\hbar$'s cancel out from Eq.\,(\ref{QP}) and the induced quantum potential is simply equal and opposite to the starting classical potential
$V$ (up to an unimportant constant
). Accordingly, the quantum force is also equal and opposite to the starting classical force. 
Second, since as mentioned above, $V$ and $V_Q$ are 
interchangeable, 
%
%
it follows that if we had started with $V_Q$ instead of $V$ in the 
Schr\"odinger equation, it would have induced an equal and opposite 
$V$ via quantum mechanics! 
In other words, what is to be treated as classical and what as quantum 
may just be a matter of choice, and the presence of {\em any} one would 
cause the other to emerge naturally, and this is true regardless of the 
exact form of the starting potential. For example, if one starts with 
an attractive gravitational potential, one would end up with a 
{\it repulsive} quantum potential. Alternatively, the observed 
attractive gravitational potential is equally likely to be, in reality, 
the quantum potential induced by a repulsive classical potential 
\cite{sd,db1,db2,perelman,dsemergent}!

Extending the idea further, we now ask a natural question: 
what if a particle experiences a potential, but the starting classical 
potential is zero, i.e. the particle is in free space, albeit with a 
wavefunction whose amplitude satisfies Eq.\,(\ref{QP})?
Evidently, the following form of Eq.\,(\ref{QP}) 
\be
{\nabla^2 \cR}  + \frac{2m\,V_Q}{\hbar^2}\,{\cal R} = 0~
\label{qp4}
\ee
%
makes it clear that the particle experiences a potential $V_Q$. However, 
there is no way to tell whether this is a background potential, or that
resulting from a specific form of the real part of the wave-amplitude 
(to be considered as time-independent and denoted by $\cR(\vx)$). 
Note that this does not completely specify the full wavefunction 
$\Psi(\vx,t)$, as its phase $\cS(\vx,t)$ remains undetermined. Nevertheless, 
the validity of the standard probability interpretation of quantum mechanics 
requires $\Psi(\vx,t)$, and hence $\cR(\vx)$, to be normalizable in the 
first place. Therefore, one can take the route of finding normalizable 
solutions of Eq.\,(\ref{qp4}) for a host of potentials. 
We however adopt a simpler approach in which $V_Q$ in Eq.\,(\ref{qp4}) is 
replaced by $V_Q + E_0$, where for simplicity, we take $E_0$ as the ground 
state energy of the system under consideration. So, the Eq.\,(\ref{qp4}) 
becomes
\be
{\nabla^2 \cR}  + \frac{2m\,\left[E_0 - (-V_Q)\right]}{\hbar^2}\,
{\cal R} = 0 \,.
\label{qp5}
\ee
%
This is nothing but the time-independent Schr\"odinger equation 
corresponding to the ground state, but for a potential $-V_Q$. 
However, the interpretation is quite different here.
Let us assume that there exist normalizable solutions of this equation, for 
some reasonable potentials. Now, the requirement of $\cR$ to be real is 
guaranteed at least for the ground state 
\cite{realpsi}.
We can therefore stipulate the potential $-V_Q-E_0$ (or approximately, 
$-V_Q$, for $|V_Q|\gg |E_0|$) to be that felt {\it effectively} by a particle, 
provided the amplitude of the particle's wavefunction, naturally arising 
or otherwise, is a solution of Eq.\,(\ref{qp5}). 

%

As an example, consider an inverted linear harmonic oscillator and 
postulate that the `observed' 
potential $V=-m\,\omega^2 x^2/2\,$ is in reality the quantum potential
experienced by a free particle, described by an wave-function whose
amplitude satisfies Eq.\,(\ref{qp5}) with $V_Q=V$, that now reads
%
\be
{\nabla^2 \cR}  +\, \frac{2m}{\hbar^2} \left(E_0 - \frac{m\omega^2\,
x^2} 2 \right){\cal R} =\, 0 \,.
\label{qp5a}
\ee
This is just the stationary state Schr\"odinger equation for a 
linear harmonic oscillator potential. The solution is of course
the unique, normalizable ground state wavefunction given by
%
%
%
\begin{eqnarray}
{\cal R} = \left(\frac{m\omega}{\pi\,\hbar}\right)^{1/4}\,
e^{- m\omega x^2/2\,\hbar} \,.
\end{eqnarray}
%
%
%
%
%
%
%
%
%
%
Similarly, substituting for a repulsive Coulomb potential $V_Q = e^2/r$,
and replacing it with $(V_Q + E_0)$ in Eq.\,(\ref{qp5}), where 
$E_0 = - me^4/(2\,\hbar^2)$ is the ground state energy of the Hydrogen atom, 
one finds the unique normalizable solution
\begin{eqnarray}
\cR =\, \frac{a_0^{-3/2}}{\sqrt{\pi}}\, e^{-r/a_0}~,
\end{eqnarray}
with $a_0 = \hbar^2/(m e^2)$.
One may carry on this way by considering any other potential, and deriving 
the corresponding $\cR$ via Eq.\,(\ref{qp5}). 
%
%
%
%
\begin{figure}[tp!]
    \centering
    \includegraphics[width=0.45\textwidth]{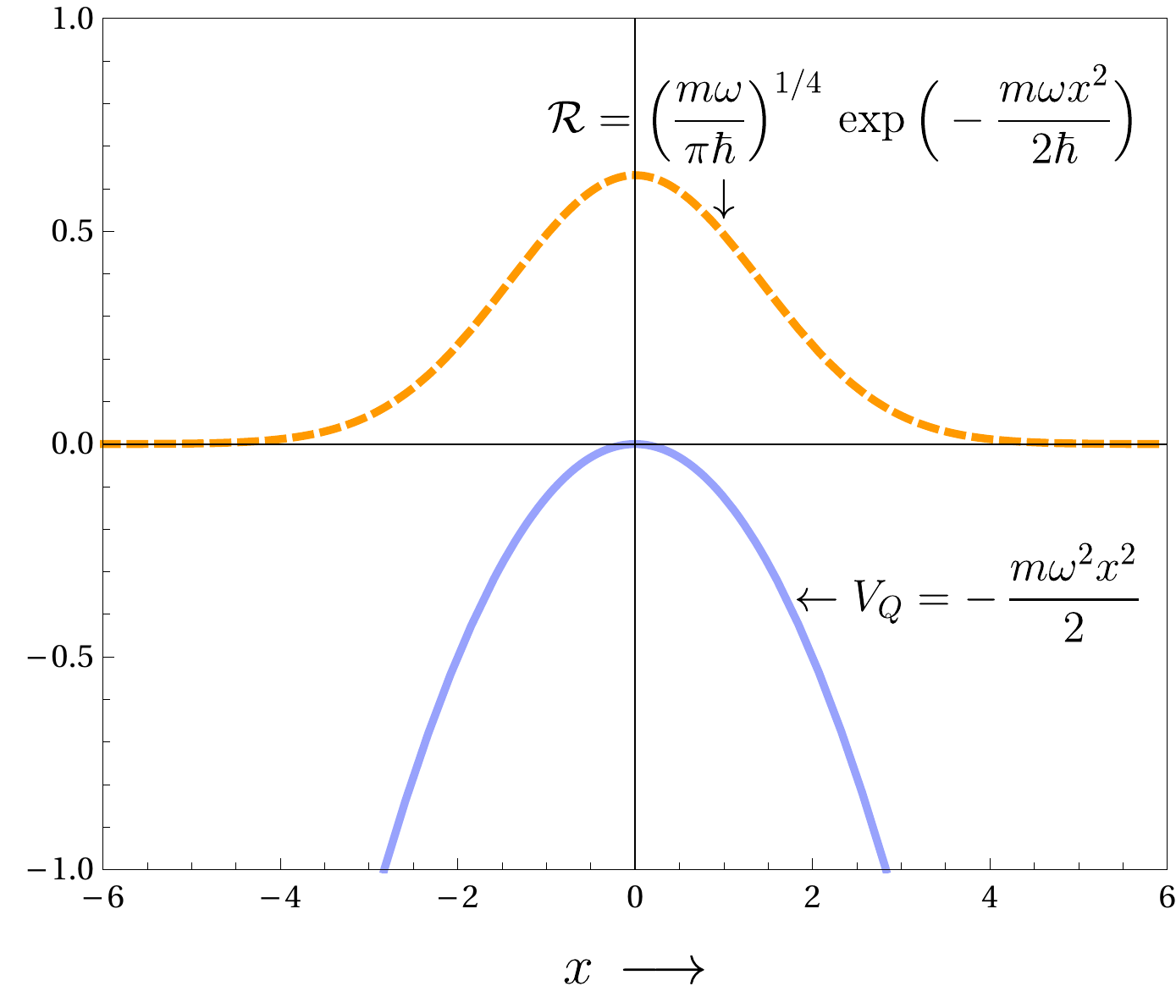}
    \caption{\small Quantum potential $V_Q=-m\omega^2 x^2/2$ 
    and its source wavefunction $\cR$, for the fiducial settings $m=1,
    \omega = 1/2$. }
    \label{fig:mesh1}
\end{figure}
\begin{figure}[tp!]
    \centering
    \includegraphics[width=0.45\textwidth]{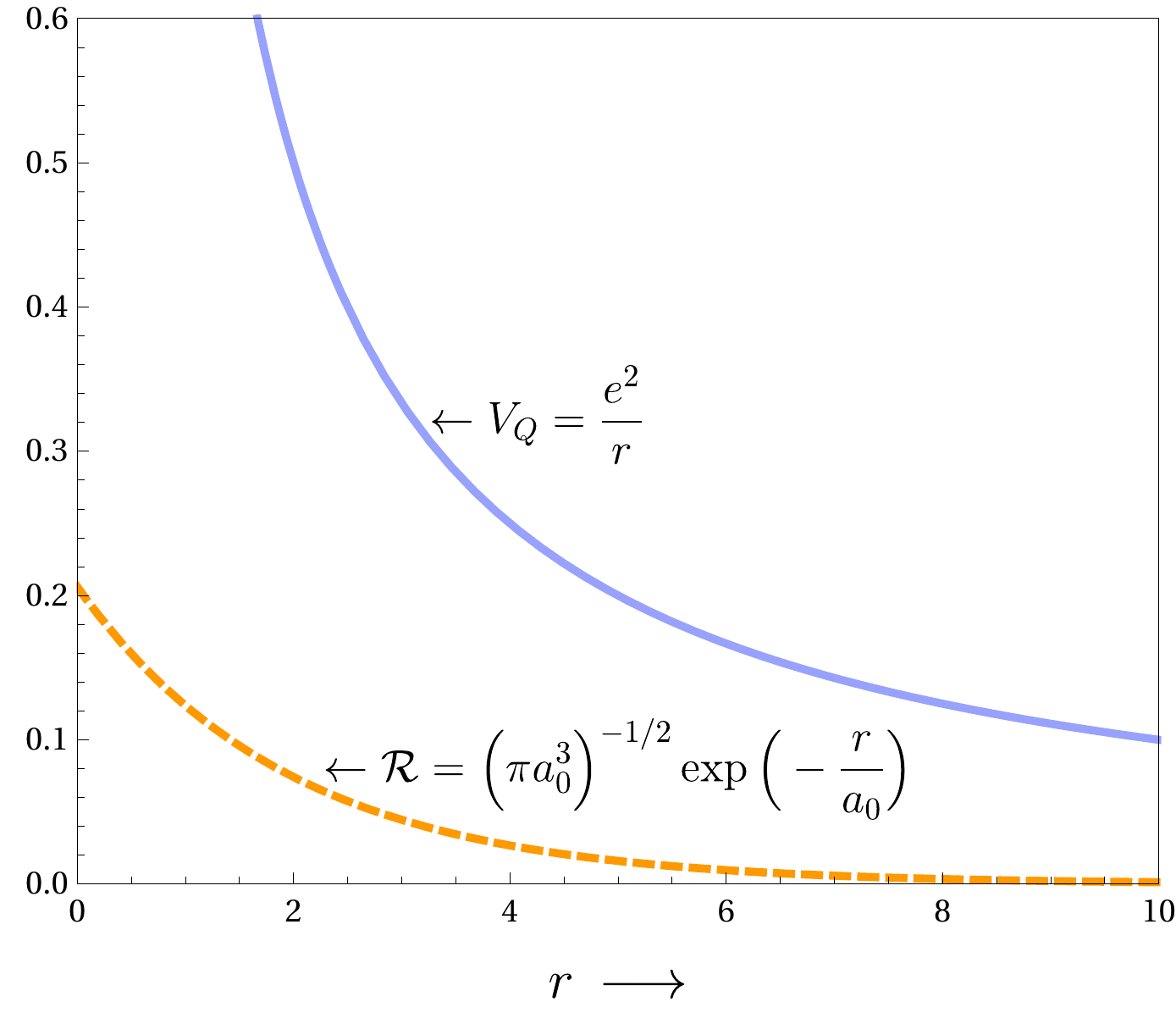}
    \caption{\small Quantum potential $V_Q=e^2/r$ and its source
    wavefunction $\cR$, for the fiducial settings $e=1, m=0.511$. }
    \label{fig:mesh2}
\end{figure}

{
To test the robustness of the method, we consider a couple of more examples and obtain the wavefunction required to produce a given quantum potential, using 
Eq.(\ref{qp4}) directly. 
First, the `step potential' (in one-dimension), defined by $V_Q=0$ for $x<0$, 
and $V_Q=V_0$ for $x\geq 0$, where $V_0$ is a constant. The corresponding 
wavefunction is given by $\cR = 1$ for $x<0$, and $\cR =\cos(kx)$ for $x\geq 0$, 
where $k= \sqrt{2m V_0/\hbar^2}$. It can be easily verified that Eq.\,(\ref{QP}) 
gives the required step potential. 
}

{
The last example we consider is a linear potential, also in one dimensions, 
given by $V_Q=\kappa x$, where $\kappa$ is a constant. In this case, 
substituting in Eq.\,(\ref{QP}) and solving for $\cR$, we get
$\cR = Ai( - k_1^{1/3}\,x)$ or $\cR = Bi( - k_1^{1/3}\,x)$, where
$Ai$ and $Bi$ are the Airy functions and $k_1=2m\kappa/\hbar^2$. 
}

{
The above examples emphasize the suitability of the method, although we note 
that the number of exactly solvable examples from Eq.\,(\ref{QP}) is limited, 
and one would have to resort to numerical techniques for more complicated 
potentials. The plots of the various quantum potentials and the corresponding 
wavefunctions that give rise to them are shown in the Figures 1-4.
}

\begin{figure}[tp!]
    \centering
    \includegraphics[width=0.45\textwidth]{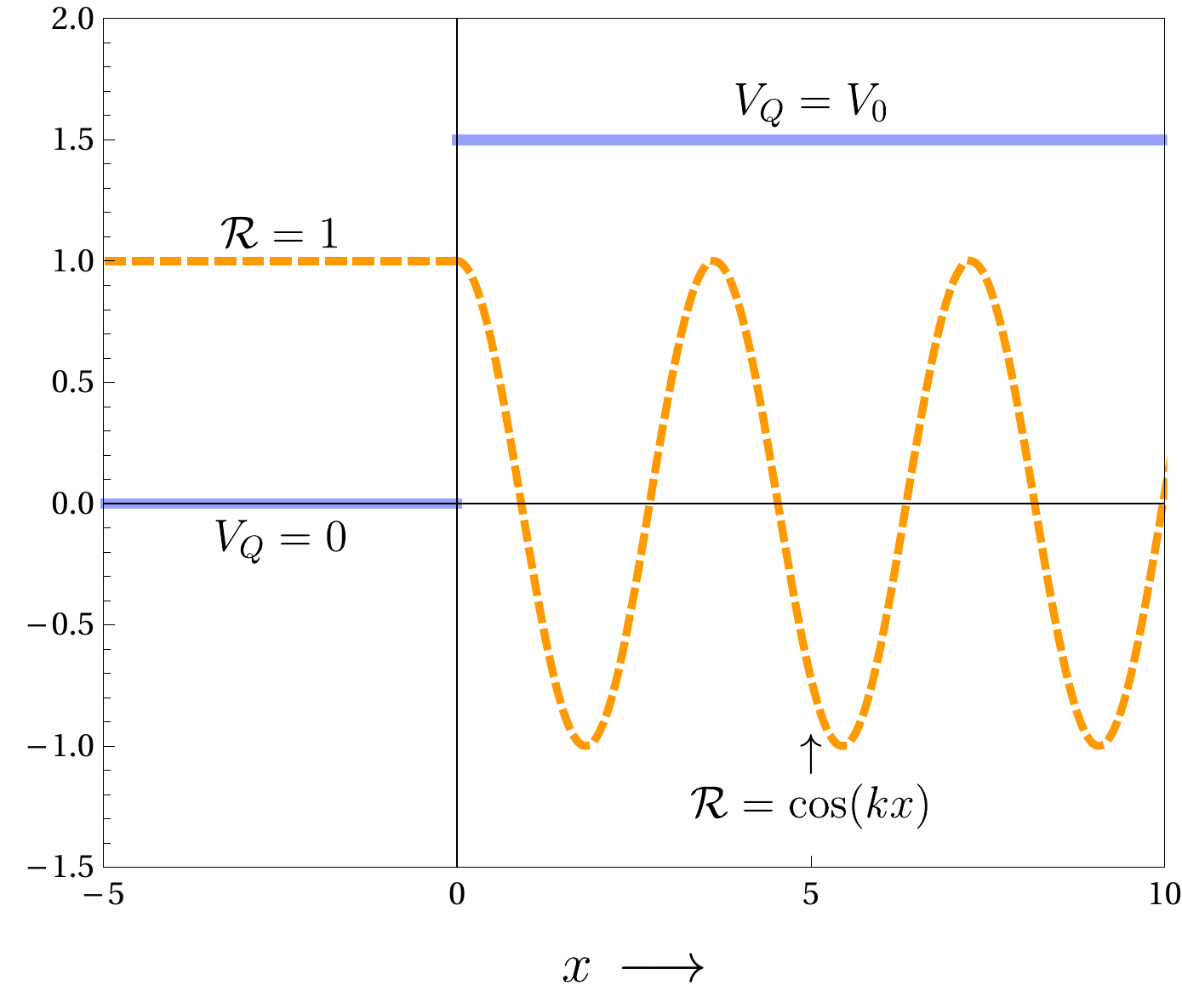}
    \caption{\small Quantum potential $V_Q=0~(x<0)$ and $V_Q = V_0
    ~(x\geq 0)$ and its source wavefunction $\cR$, for the fiducial 
    settings $m=1, V_0=1.5$.}
    \label{fig:mesh3}
\end{figure}
\begin{figure}[tp!]
    \centering
    \includegraphics[width=0.45\textwidth]{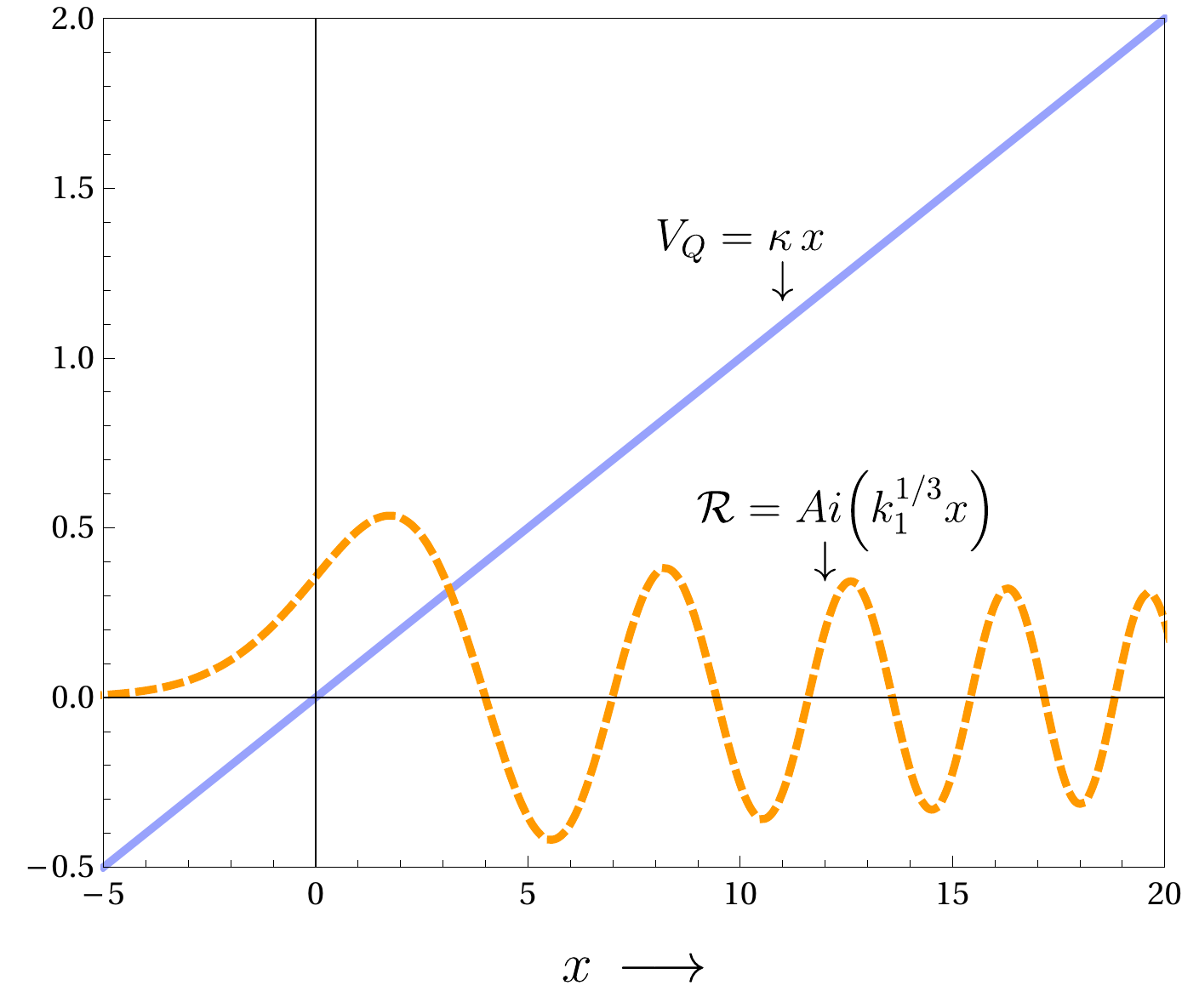}
    \caption{\small Quantum potential $V_Q=\kappa\,x$  and its source 
    wavefunction $\cR$, for the fiducial settings $m=1, \kappa=0.1$.}
    \label{fig:mesh4}
\end{figure}

To summarize, we have shown that practically any presumptive classical 
potential can in fact be derived as the quantum potential of a suitable 
wavefunction. These could be familiar potentials such as the Coulomb,
inverted harmonic oscillator or the gravitational potential, or exotic 
ones such as those which can effectively explain the flat galaxy rotation 
curves. Therefore it begs the questions as to whether the ones we observe 
as potentials and interactions are mere manifestations of appropriate 
wavefunctions. The question is particularly important in the context of 
the gravitational field, since the corresponding field theory is known 
to be non-renormalizable. Therefore, a more fundamental picture, with 
the observed gravitational fields as quantum potentials of particle 
wavefunctions would offer a fresh insight into the problem. It should 
even predict corrections to the $1/r$ potential, at very small or large 
length scales, beyond the regimes in which the Newtonian gravity has 
been verified experimentally. 
Of course, there are many questions to be answered. Some of these should 
be straightforward, for example, the relativistic generalization of our 
work starting from a relativistic wave equation (part of it has been done 
in ref.\,\cite{sd}). The others, for instance, the determination of the 
phase $\cS$ and how the appropriate wavefunctions can be specified, 
require further studies. 

To conclude, we list a few important observations and possible directions
for future works. 

\vspace{3pt}
\noindent
{
1. {\bf Excited states:}
we remind that Eq.\,(\ref{qp5}) is an approximation of Eq.\,(\ref{qp4}), 
where the ground state energy $E_0$ was introduced in an attempt to 
guarantee a normalizable $\cR$, albeit at the cost of `raising' the 
quantum potential by a small amount. Similarly, one may replace 
$E_0 \rightarrow E_n$, an excited state energy, and $\cR \rightarrow \cR_n$, 
the corresponding excited state. This demonstrates the non-uniqueness of the 
wavefunction in general for a given quantum potential.  
}

\vspace{3pt}
\noindent
{
2. {\bf Non-stationary states:} as Eq.\,(\ref{qp4}) suggests, a time-dependent 
quantum potential $V_Q$ would require a time-dependent $\cR$ as its source. 
Furthermore, one sees from Eq.\,(\ref{consv-eq}) that such a $\cR$ would entail 
a non-trivial phase $S$. 
}

\vspace{3pt}
\noindent
{
3. {\bf Other formulations:}
It may be noted that a `modified potential', distinct from, but related to 
the quantum potential, was defined and its implications explored in detail in
\cite{floyd}. A re-examination of our results in light of the aforementioned 
potential may shed further light on emergent potentials. Similarly, the possible 
origin of the Schr\"odinger equation purely from symmetry considerations was 
examined in \cite{faraggi1,faraggi2}. That study may also complement the current 
work, as it provides a justification for starting with the Schr\"odinger equation 
in the first place. It may also be useful in extending our results to other 
quantum wave equations. 
}

\vspace{3pt}
\noindent
{
4. {\bf Future directions:} in addition to studying the above issues, namely 
excited states, non-stationary states and other formalisms, we will carefully 
investigate a few other things. This includes, the (special) relativistic 
generalization of our results, computing corrections to the gravitational 
potential at various length scales, studying the stability of the wavefunction, 
generalization to higher dimensions and last but not the least, examine whether 
the idea can be generalized to Yang-Mills potential. We hope to report on these 
elsewhere. 
}


\vspace{5pt}
\noindent 
{\bf Acknowledgment}

\no
This work was supported by the Natural Sciences and Engineering Research Council 
of Canada. We thank the anonymous referee for useful comments which have helped us 
to improve the manuscript. 

\end{document}